\documentclass[a4paper,12pt]{article}
\usepackage{times}
\usepackage[final]{graphicx}            
\usepackage{citesort}

\begin{document}

\title{\bfseries \Large
Terminating ventricular tachycardia
by pacing induced dynamical inhomogeneities in the reentry circuit
\vspace*{-12pt}
}
\author{ \large
Sitabhra Sinha\footnote{E-mail: sitabhra@imsc.res.in} ~
and Johannes Breuer\footnote{Also at: Institute of Theoretical Physics,
Technical University-Berlin, D-10623 Berlin, Germany.}
}
\date{
\vspace*{-8pt}
\normalsize
The Institute of Mathematical Sciences, C. I. T. Campus,
Taramani, Chennai - 600 113, India.
}
\maketitle
\thispagestyle{empty}
\normalsize

\emph{Abstract-}\textbf{
Formation of feedback loops of excitation waves (reentrant
circuit) around non-conducting ventricular scar tissue is a common cause
of lethal cardiac arrhythmias, such as ventricular tachycardia. 
This is typically treated by rapid stimulation
from an implantable device (ICD). However, the mechanisms of reentry
termination success and, more importantly, failure, are poorly understood. 
To study such mechanisms, we simulated pacing of anatomical reentry in
an ionic model of cardiac tissue, having significant restitution and dispersion
properties. Our results show that rapid pacing causes inhomogeneous
conduction in the reentrant circuit, leading to
successful pacing termination of tachycardia.
The study suggests that more effective pacing algorithms can be
designed by taking into account the role of such dynamical inhomogeneities.}

\begin{center}
\textbf{I. I\textsc{ntroduction}}
\end{center}
The heart is an extremely efficient mechanism that mediates the
transmission of deoxygenated blood to the lungs and oxygenated
blood to the rest of the body by rhythmic contraction of its
two upper chambers (atria) followed by the
two lower chambers (ventricles). The mechanical action of contraction,
and subsequent relaxation, is initiated by electrical activity in
the excitable cells of the heart wall. In the resting state, cardiac cells
are in a hyperpolarized state with a membrane potential 
of $\sim -85$ mV. However, upon being excited by a sufficiently large
stimulus (i.e., a stimulus which is sufficient to increase the membrane
potential beyond the excitation threshold, $\sim -60$ mV), 
they are rapidly depolarized
to a membrane potential of $\sim 30$ mV, followed by a plateau phase
when the membrane potential remains at a steady value for some time,
and then ultimately coming back to the resting state following an extended
period of slow repolarization. 
This series of steps constitute the action potential (AP) and all the 
processes in it are mediated by the action of voltage-sensitive
ion channels located on the cell membrane that are selective for different
charged ions, such as, Na$^+$, K$^+$ and Ca$^{++}$. In human ventricles,
the duration of the AP is $\sim$ 200 msec. Neighboring cells are connected
to each other by gap junctions, which allows the excitation to propagate
from cell to cell through currents due to differences in their 
membrane potential. Waves of excitation are therefore observed 
to propagate along the heart wall. Two such waves annihilate each other
upon collision because of the existence of a refractory period in 
cardiac cells. Refractory period refers to the time during which the 
cardiac cell slowly recovers its resting state properties after an episode of  
activation. During this time the cell cannot be excited even if 
a suprathreshold stimulus is applied. 

During normal functioning of the heart, the sinus node acts as 
the natural oscillator dominating the rhythm of activation. The excitation
propagates throughout the atria from the sinus node. As the ventricles are
electrically isolated from the atria, the propagation of excitation
to the lower chambers of the heart can occur only via the slow conduction
pathway of the AV node. This serves the dual purpose of allowing a time delay
between the activation of the atria and the ventricles (thus allowing
the blood from the atria to be fully pumped into the ventricles, before
contracting the latter), as well as, protecting the ventricle from being
affected by very rapid excitations in the atria. The excitation then 
propagates throughout the ventricles, from the apex to the base.
Under certain situations (e.g., in people suffering from an ischemic heart), 
this normal activity can be hampered by arrhythmias,
i.e., disturbances in the natural rhythmic activity of the heart \cite{Win87}.
A potentially fatal arrhythmia occurring in the ventricles is tachycardia, or
abnormally fast excitation, during which the heart can beat as rapidly as
300 beats per minute. There are multiple mechanisms by which ventricular
tachycardia (VT) may arise, but the most common one is due to the
formation of a reentrant pathway, i.e., a closed path of excitation feedback.
Reentry can have an anatomical substrate, with the excitation wave going round
and round an existing inexcitable obstacle, e.g., a region of
scar tissue as shown in Fig. 1~(left). 
Reentry can also occur around a region
which was only transiently inexcitable (e.g., during recovery from a 
premature excitation), and once established, will persist
even when the region has recovered its excitability. However, in this
paper, we will focus on reentry around an anatomical obstacle.

For people in chronic risk of VT, the most common treatment is implanting
an ICD, a device capable of detecting the onset of VT and giving a sequence of
low-amplitude electrical pulses (pacing) through an electrode, usually
located in the ventricular apex, in order to restore the normal 
functioning of the heart \cite{Jos93}. 
The operating principle of this device is
that by applying pacing at a frequency higher than that of the VT,
the pacing waves will eventually reach the reentrant circuit and 
terminate the reentry. 
However the underlying mechanisms of the success and
failure of pacing termination are not yet well-understood
and the algorithms currently used in such devices are often based on
purely heuristic principles. As a result, occasionally, 
instead of terminating VT,
pacing can accelerate it or can even promote its degeneration to 
lethal ventricular fibrillation (VF), leading to death within minutes if
no immediate action is taken. Understanding the interaction between 
pacing and reentrant waves is therefore essential for designing more effective
and safer ICD pacing algorithms.

For ease of theoretical analysis and numerical computations, most studies
of pacing have focussed on reentry in a one-dimensional ring of cardiac
cells, which is essentially the region immediately surrounding an anatomical
obstacle \cite{Gla95,Nom96,Com02,Sin02a,Sin02b}. The conventional
view of reentry termination has been that each pacing wave splits
into two branches in the reentry circuit, the retrograde branch traveling
opposite to the reentrant wave and eventually colliding with it, 
annihilating each other. The other, anterograde, branch travels in
the same direction as the reentrant wave, and depending on the timing
of the pacing stimulation, either resets the reentry by becoming the new
reentrant wave, or leads to termination, if it is blocked by a 
refractory region left behind in the wake of the preceding wave.
If the pacing site is on the ring itself, continuity arguments can
be used to show that there will always exist a range of stimulation
times, such that the reentry will be terminated. However, this argument
breaks down when we consider a pacing site situated some distance away from
the reentry circuit. As, in reality, the pacing site is fixed (usually in
the ventricular apex), while the reentry can occur anywhere in the
ventricles, this leaves the question open about how pacing terminates
VT.

\begin{figure}[t!]
	\centerline{\includegraphics[width=0.85\linewidth,clip] {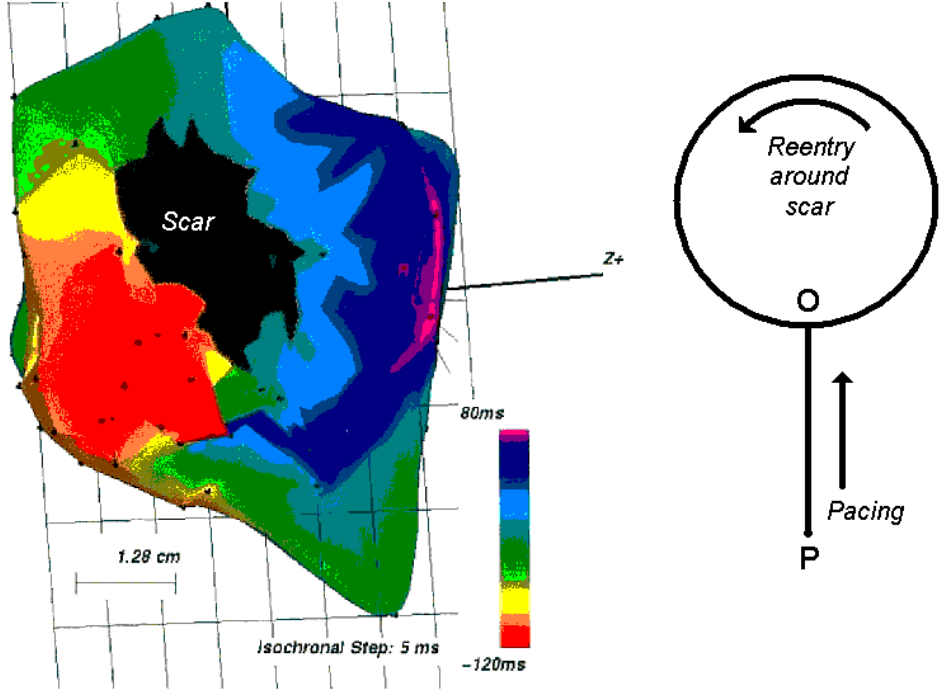}}
	\vspace*{-8pt}
	\caption{\footnotesize (Left) Electrophysiological activity map 
	during tachycardia in a
	human ventricle obtained using the Biosense-Webster CARTO EP 
	Mapping System (Weill Medical College of Cornell University, New York).
	The colors code for excitation occurring at different
	times. Note the non-conducting scar tissue (in black) occupying a 
	significant portion of the ventricle. Anti-tachycardia pacing
	is usually applied via an electrode placed at the ventricular apex
	(the lowermost point of the ventricle in the figure). 
	(Right) A schematic diagram of anti-tachycardia pacing. Reentrant
	activity occurring around a scar tissue is simplified into
	a wave going around a ring. The sidebranch joining
	the ring at O represents the external stimulation arriving from the
	pacing electrode located at P.}
	\label{fig:fig1}
\end{figure}
To address this issue, we have previously investigated reentry in a 
quasi-1D geometry consisting of a ring attached to a sidebranch, the
other end of which is the pacing site (Fig.~1, right). Our studies
showed that existence of inhomogeneities in the reentry circuit
are essential for successful termination of VT by pacing \cite{Sin02b}.
Studies in two dimensions upheld the qualitative results \cite{Sin02a}.
However, we had focussed exclusively on the role of anatomical inhomogeneities, 
such as a zone of slow conduction. 

To simplify
analysis, we had considered long reentrant circuits where restitution and
dispersion effects of cardiac tissue can be neglected. These two 
effects, where the duration of the action potential and the conduction
velocity, respectively, of an excitation wave, depend on the
time interval from the preceding wave, have recently been shown
to cause inhomogeneities in the properties of cardiac tissue \cite{Ech02}, 
sometimes leading to conduction block \cite{Fox02}. In this paper, we
return to the problem of reentry termination, and show that the
generation of dynamical heterogeneities, by pacing in shorter 
reentrant circuits, can lead to successful VT termination.
We believe this to be the principal mechanism by which ICDs terminate
VT. Our study, therefore, has implications for designing effective 
pacing algorithms.

\begin{center}
\textbf{II. M\textsc{odel}}
\end{center}
As mentioned before, we consider a quasi-1D geometry consisting of a
ring of model cardiac cells, attached to a sidebranch. 
The propagation of excitation in this model is described by the partial
differential equation:
\begin{equation}
{\partial V}/{\partial t} = I_{ion} / C_m + D {\nabla}^2 V,
\end{equation}
where $V$ (mV) is the membrane potential, $C_m$ = 1 $\mu$F cm$^{-2}$ is
the membrane capacitance, $D$ (cm$^2$~sec$^{-1}$) is the diffusion
constant and $I_{ion}$ ($\mu$A cm$^{-2}$) is the cellular membrane
ionic current density. We used the Luo-Rudy I action potential model
\cite{Luo91}, in which $I_{ion}$ = $I_{Na}$ + $I_{si}$ + $I_{K}$
+ $I_{K1}$ + $I_{Kp}$ + $I_b$. $I_{Na} = G_{Na}m^3hj(V - E_{Na})$
is the fast inward Na$^+$ current, $I_{si} = G_{si}df(V - E_{si})$
is the slow inward Ca$^{++}$ current, $I_{K} = G_{K}xx_i(V-E_{K})$ is the
slow outward time-dependent K$^+$ current, $I_{K1} =
G_{K1}K1_{\infty}(V-E_{K1})$ is the time-independent K$^+$ current,
$I_{Kp} = 0.0183 K_p(V-E_{Kp})$ is the plateau K$^+$ current, and
$I_b = 0.03921(V+59.87)$ is the total background current.
$m$, $h$, $j$, $d$, $f$, $x$ and $x_i$ are the gating variables satisfying
differential equations of the type: $dy/dt = (y_{\infty}-y)/{\tau}_y$,
where $y_{\infty}$ and ${\tau}_y$ are dimensionless quantities which are
functions solely of $V$. The external K$^+$ concentration is set to be
$[$K$]_0$ = 5.4mM, while the intracellular Ca$^{2+}$ concentration
obeys $d [{\rm Ca}]_i/dt =
-10^{-4} I_{si} + 0.07(10^{-4}- [{\rm Ca}]_i)$.
The details of the expressions and the values used for the constants
can be found in Ref.\cite{Luo91}. However, in accordance with Ref. \cite{Qu99},
two of the conductance parameters,
$G_{si}$ and $G_K$ have been changed from their original values 
to 0.07 mS cm$^{-2}$ and 0.705 mS cm$^{-2}$, respectively,
to shorten the duration of the action potential to that observed
in the human ventricle. We choose $D$ = 0.556 cm$^2$~sec$^{-1}$, so that
the conduction speed of an excitation wave is $\sim$ 47 cm~sec$^{-1}$ in
fully recovered tissue.

We solve the model  by using a forward-Euler integration scheme. We
discretise the system on a grid of points in space with spacing
$\delta x = 0.01\/$ cm and use the standard 3-point
difference stencil for the 1-D Laplacian, except at the junction of
the ring and sidebranch, where we used a 5-point stencil for 2-D Laplacian.
The spatial grid consists of a linear lattice with $L$ points for the
ring and $L^{\prime}$ points for the sidebranch;
in this study we have used values of $L$ = 900 and $L^{\prime}$ = 300.
The time step is $\delta t = 0.005$ msec.
Each pacing pulse is 28 $\mu$A cm$^{-2}$ and applied for 2 msec, which is 
just sufficient to elicit an action potential
in fully recovered cardiac tissue.
\begin{center}
\textbf{III. R\textsc{esults and Discussion}}
\end{center}
We initiate reentry in the model by a two stimulus protocol, with the
second stimulus applied at a point that has just recovered from excitation
due to the first stimulus, so that the resultant wave can propagate in
a single direction only. We next allow the reentrant propagation to
attain a steady state, completing 24 turns around the ring. The ring
length being short enough for restitution and dispersion effects to
be significant, the steady state shows discordant alternans, with 
succeeding waves at a particular location on the circuit alternating 
between long and short durations of action potential, as well as, slightly 
different conduction velocities. E.g., at the point O, the duration of
the action potential of succeeding waves alternates between $\sim$148 msec
and $\sim$59 msec, and the corresponding conduction velocities are
46.5 and 45.7 cm~s$^{-1}$. The reentry period is, on the average, 195 msec.

Next, we observe the effect of pacing, using a sequence of 8 pulses with
a constant time interval ($PI$) between them. This is known as {\em burst
pacing} in the clinical literature \cite{Hor95}, where usually 
6$-$10 pulses are delivered at a constant frequency. 
In the other commonly used pacing protocol,
{\em ramp pacing}, the time interval between pulses is gradually
decreased over the
course of pacing. However, our preliminary simulations 
show little difference in termination performance
between burst and  ramp pacing. The pacing period is usually between 
80-90$\%$ of the reentry period, and for our simulations,
$PI$ is scanned through this range.
Besides $PI$, the other pacing parameter is the timing of the initial
pulse, measured by coupling interval ($CI$), the time interval between
the activation of the pacing site by the preceding reentrant wave
and the first pacing pulse.
\begin{figure}[t!]
        \centerline{\includegraphics[width=0.85\linewidth,clip] {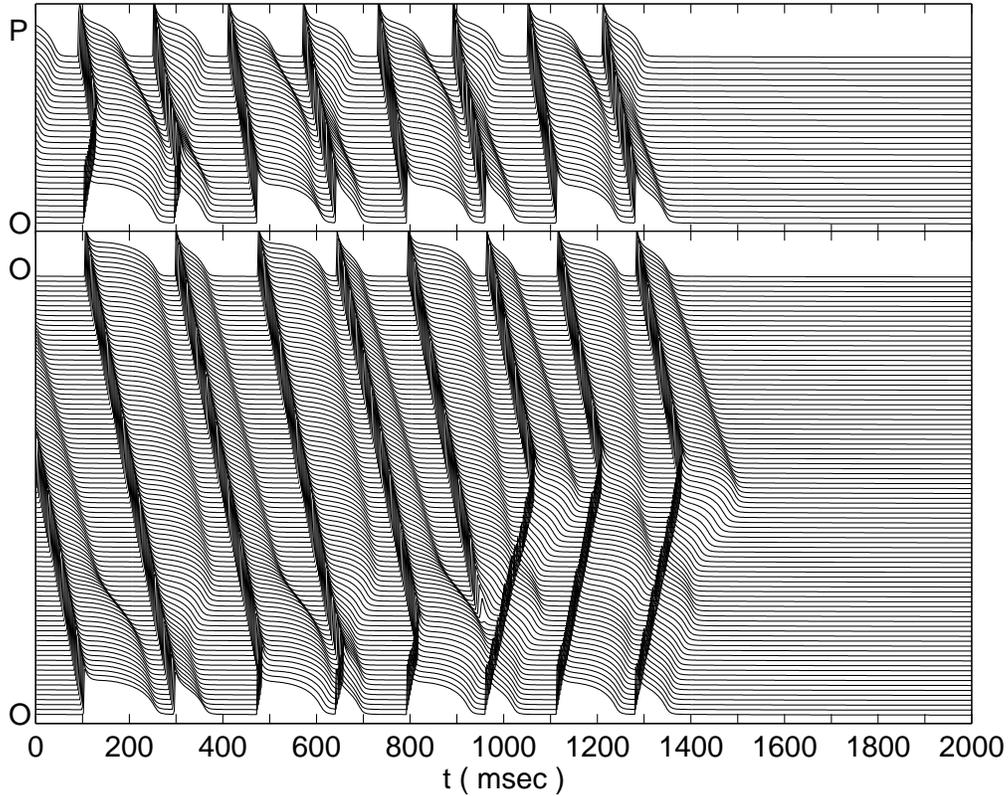}}
        \vspace*{-8pt}
        \caption{\footnotesize Space-time plot of membrane potential ($V$)
	showing pacing termination of reentrant wave, having a
	period of $\sim$ 195 msec, in an arrangement
	consisting of a sidebranch (top) and ring (bottom) of cardiac cells 
	modeled using the Luo-Rudy I equations. The ring is of length 9 cm,
	while the pacing electrode (at P) is located 3 cm away from the ring.
	The junction point of the ring and sidebranch is indicated by O.
	Eight pacing pulses having a period of 160 msec ($\sim 82 \%$ of the
	reentry period) are applied at P
	starting at $t \simeq$ 86 msec. The pacing wave enters the
	ring after the third pulse and termination of reentry occurs after
	the fifth pacing pulse, around $t$ = 954 msec
	at a distance of 1.85 cm from O.}
        \label{fig:fig2}
\end{figure}

Fig. 2 shows an instance of successful termination of reentry for
a pacing cycle of $PI$ = 160 ms, which is roughly 82$\%$ of the reentry 
period, and $CI$ = 115.16 msec. To understand how pacing terminates VT,
note that the higher frequency pacing wave enters the reentry circuit
after the third pulse and perturbs the established steady state. The
reentry period decreases suddenly from 195 msec, and wave propagation
in the ring is destabilized as the conduction velocity tries to adjust to this
lower cycle. As a result, a localized region of slow conduction is 
produced in the region neighboring the point where the pacing wave
collides with the reentrant wave. As detailed in our previous study
\cite{Sin02b}, the slow conduction allows smaller and smaller amount of
electrical conduction between neighboring cells, until, at the fifth
pacing beat, the current is insufficient to initiate an action potential
in the cell immediately in front of the excitation wave. This leads
to conduction block of the anterograde branch of the pacing wave,
and therefore, to termination of reentry.

Based on the simulation results, we arrive at the following conclusions about
the impact of pacing parameters on VT termination. 
The number of pacing pulses is close to the optimal, 
as using larger number of pulses often cause further conduction blocks 
and, as a result, restarts the reentry. As the number of pulses required
for termination
is dependent on the distance between the pacing site and the reentrant 
circuit, the upper limit on the number can be related to the spatial
extension of the ventricles.
From this it follows that the location of the pacing site, relative to
the ring, often decides whether a given pacing protocol
will succeed in terminating reentry.

The pacing frequency has to be carefully chosen. While $PI$ has to be
shorter than the VT period, it cannot be too short, as the propagation
of high frequency waves causes instability and wave breakup, leading to
formation of spiral waves around transiently inactive
cores (`functional' reentry). This maybe the mechanism
responsible for rapid pacing occasionally giving rise to faster arrhythmias.
Wave instability can initiate further breakup of the spiral wave leading
to the spatiotemporal chaos of VF.
The timing of the initial pulse is also crucial. If the coupling
interval is small, the pacing wave is blocked by the refractory
region left behind by the preceding wave, whereas if it is large, the
pacing wave will collide with the next reentrant wave further and further
from the ring and may not be able to enter the reentrant
circuit at all.
The amplitude of stimulation also plays a very important role. 
By increasing the pacing amplitude 
to 40 $\mu$A~cm$^{-2}$, we have significantly increased
the range of coupling intervals over which successful reentry termination is
achieved.

To verify the model independence of our conclusions we have also looked at the
much simpler Karma model \cite{Kar94} and found qualitatively similar 
results. We are currently working on extending our analysis to 2-D
and 3-D simulations of cardiac tissue.
The ultimate goal of anti-tachycardia pacing is to terminate reentrant
activity with pulses of smallest magnitude in the shortest possible time
with the lowest probability of giving rise to faster arrhythmias or
VF. The constant frequency pacing investigated here is only a partial
solution to this end, and a more efficient algorithm might have to
adjust the pacing intervals on a beat-to-beat basis.
The results of our investigation is aimed towards answering how
such an optimized pacing scheme maybe designed.

\begin{center}
\textbf{A\textsc{cknowledgment}}
\end{center}
We thank David Christini for helpful discussions.
J.B. would like to thank DAAD for financial support. 

\vspace{2pt}

\begin{center}
\textbf{R\textsc{eferences}}
\end{center}

\end{document}